%% file: main.tex
\documentclass[APS,Times2COL]{WileyNJDv5} 

\usepackage{bm}

\articletype{Research Article}%

\received{Date Month Year}
\revised{Date Month Year}
\accepted{Date Month Year}
\journal{Journal}
\volume{00}
\copyyear{2023}
\startpage{1}

\newcommand{\velocityField}{\mathbf{u}}

\newcommand{\forceBuoyancy}{\mathbf{f}_{b}}
\newcommand{\forceExternal}{\mathbf{f}_{e}}
\newcommand{\gravity}{\mathbf{g}}
\newcommand{\positionCoordinate}[1]{\mathbf{X}_{{#1}}}
\newcommand{\dropletVelocity}[1]{\mathbf{V}_{{#1}}}

\newcommand{\review}[1]{#1}

\begin{document}

\title{Towards Accelerating Particle-Resolved Direct Numerical Simulation with Neural Operators}

\author[1]{Mohammad Atif}
\author[1,2]{Vanessa L{\'o}pez-Marrero}
\author[3]{Tao Zhang}
\author[4]{Abdullah Al Muti Sharfuddin}
\author[1]{Kwangmin Yu}
\author[5]{Jiaqi Yang}
\author[3]{Fan Yang}
\author[4]{Foluso Ladeinde}
\author[3]{Yangang Liu}
\author[1]{Meifeng Lin}
\author[1]{Lingda Li}

\authormark{ATIF \textsc{et al.}}
\titlemark{Towards Accelerating Particle-Resolved Direct Numerical Simulation with Neural Operators}

\address[1]{\orgdiv{Computational Science Initiative}, \orgname{Brookhaven National Laboratory, Upton}, \orgaddress{\state{New York 11973}, \country{USA}}}

\address[2]{\orgdiv{IACS}, \orgname{Stony Brook University }, \orgaddress{\state{New York}, \country{USA}}}

\address[3]{\orgdiv{Environmental and Climate Sciences Department}, \orgname{Brookhaven National Laboratory, Upton}, \orgaddress{\state{New York 11973}, \country{USA}}}

\address[4]{\orgdiv{Department of Mechanical Engineering}, \orgname{Stony Brook University }, \orgaddress{\state{New York}, \country{USA}}}

\address[5]{Department of Mathematics, Emory University, Atlanta, GA 30322, USA}

\corres{Mohammad Atif, \email{fmohammad@bnl.gov}}



\abstract[Abstract]{    
We present our ongoing work aimed at accelerating a particle-resolved direct numerical simulation model designed to study aerosol-cloud-turbulence interactions. The dynamical model consists of two main components -- a set of fluid dynamics equations for air velocity, temperature, and humidity, coupled with a set of equations for particle (i.e., cloud droplet) tracing. Rather than attempting to replace the original numerical solution method in its entirety with a machine learning (ML) method, we consider developing a hybrid approach. We exploit the potential of neural operator learning to yield fast and accurate surrogate models and, in this study, develop such surrogates for the velocity and vorticity fields. 
We discuss results from numerical experiments designed to assess the performance of ML architectures under consideration as well as their suitability for capturing the behavior of relevant dynamical systems.
}



\maketitle

\renewcommand\thefootnote{}

\renewcommand\thefootnote{\fnsymbol{footnote}}





\section{Introduction}  \label{sec:intro} 
Particle-resolved direct numerical simulations (PR-DNS), which resolve not only the smallest turbulent eddies but also track the development and motion of individual particles, are essential tools for studying aerosol-cloud-turbulence interactions \cite{gao2018investigation,cloud_modeling_workshop}.  The associated mathematical models comprise systems of partial differential equations (PDEs), including the Navier-Stokes equations for fluid flow.  In turbulent regimes, the numerical solution of the Navier-Stokes equations comes at a high computational cost.  To reduce the cost of such high-fidelity direct numerical simulations (DNS), surrogate models (of lower fidelity) are often employed.  With judicious algorithmic choices, high-fidelity models may be used in conjunction with surrogate models to reduce computing costs, while retaining acceptable levels of accuracy in the simulation results \cite{review_multifidelity_methods,Spyropoulos_2022}.

The confluence of machine learning and computational fluid dynamics has led to important progress in increasing the speed of DNS \cite{kochkov2021machine}, turbulence modeling \cite{duraisamy2019turbulence}, developing surrogate models \cite{kutz2016dynamic}, and learning patterns and PDEs from data \cite{brunton2016discovering,tucny2023learning,wang2022finding}.
Among many recent advancements in machine learning (ML) and artificial intelligence, deep neural networks for operator learning, which learn mappings between function spaces \cite{chen_1995, li2020fourier, deeponet_2021, JMLR:v24:21-1524}, have emerged as new scientific computing tools with much potential to aid in the development of fast and accurate surrogate models for dynamical systems \cite{li2020fourier,LU2022114778}.  Research with such neural operator networks is currently very active in many application areas, including climate sciences\cite{jiang2021DigitalTwinEarth,pathak2022fourcastnet}.  

In the present study, we consider the use of such neural operator methods in order to assess their performance as surrogates of a numerical PR-DNS model.  Due to the complexity of the PR-DNS model of interest to us, as a first step we focus on a subset of equations used in the PR-DNS model rather than attempting to replace the physical model in its entirety with a machine learning surrogate.  This will allow us to study the potential benefits of a hybrid approach, where the neural operator surrogates are judiciously incorporated within the numerical PR-DNS model. 

We proceed to describe the PR-DNS model in Section \ref{sec:model_equations}.  Our initial test case and data set are presented in Section \ref{sec:dataset}.  The methodology and choice of neural operator method appears in Section \ref{sec:methodology}, followed by a discussion of the obtained results in Section \ref{sec:results}.  Concluding remarks and future research directions are given in Section \ref{sec:conclusion}.

\section{Particle-Resolved DNS Model}
\label{sec:model_equations}
A summary of the physical model ultimately of interest to us appears next.  Readers may refer to \cite{gao2018investigation} for a thorough description of the model.

\subsection{Fluid Dynamics Equations}
\label{sec:equations_u_T_q}

The incompressible Boussinesq system
\begin{eqnarray}
    \frac{\partial \velocityField}{\partial t}  + (\velocityField \cdot \nabla)\velocityField  
    & = & 
    - \frac{1}{\rho_{0}} \nabla p + \nu \nabla^{2} \velocityField + \forceBuoyancy + \forceExternal  \label{eq:NS}  \\
    \nabla \cdot \velocityField  & = &  0    \label{eq:incompressible}
\end{eqnarray} 
is adopted to model the fluid velocity.  In equations \eqref{eq:NS}--\eqref{eq:incompressible}, $\velocityField$ is the fluid velocity field, $p$ denotes pressure, $\nu$ is the kinematic viscosity, and $\rho_{0}$ is the density of dry air.  
The buoyancy force $\forceBuoyancy$ in \eqref{eq:NS} is given by
\begin{eqnarray}    
   \forceBuoyancy  & = &  - \gravity \left [ \frac{T - T_{0}}{T_{0}} + 0.608(q_{v} - q_{v_{0}}) - q_{c} \right ],   \label{eq:buoyancy_force}
\end{eqnarray}
where $\gravity$ denotes the gravitational acceleration vector, $T$ is temperature, $q_{v}$ is the water vapor mixing ratio, and $q_{c}$ is the liquid water mixing ratio.  $T_{0}$ and $q_{v_{0}}$ denote reference values.
As the buoyancy force $\forceBuoyancy$ in \eqref{eq:buoyancy_force} depends on the temperature $T$ and the water vapor mixing ratio $q_{v}$, it couples the equations \eqref{eq:NS}--\eqref{eq:incompressible} for fluid velocity with those below for temperature \eqref{eq:temperature} and water vapor \eqref{eq:vapor_mixing_ratio}.  Finally, the external force $\forceExternal$ in \eqref{eq:NS} is introduced to maintain a statistically stationary homogeneous turbulence \cite{gao2018investigation}.

The time evolution for the temperature $T$ and water vapor mixing ratio $q_{v}$ is modeled by the equations
\begin{eqnarray}    
    \frac{\partial T}{\partial t}  + (\velocityField \cdot \nabla) T  
    & = & 
    \frac{L_{h}}{c_{{p}}} C_{d}  + \mu_{T} \nabla^{2} T \label{eq:temperature}  \\
    \frac{\partial q_{v}}{\partial t}  + (\velocityField \cdot \nabla) q_{v}
    & = & 
    - C_{d}  + \mu_{v} \nabla^{2} q_{v}, \label{eq:vapor_mixing_ratio}
\end{eqnarray}
where $L_{h}$ is the latent heat of water vapor condensation, $c_{p}$ is the speciﬁc heat at constant pressure. $\mu_{T}$ and $\mu_{v}$ are the molecular diffusivities for temperature and water vapor, respectively.   The rate of exchange between liquid and vapor, a.k.a. the condensation rate $C_{d}$, as defined in equation \eqref{eq:condensation_rate}, depends on droplet radii and thus couples the fluid dynamics model with the equations for droplet growth and motion depicted in Section \ref{sec:equations_droplets} below.

\subsection{Droplet Growth and Motion}
\label{sec:equations_droplets}

The equations describing the condensation/evaporation of cloud droplets and their motion are given by
\begin{eqnarray}    
    R_{i}(t) \, \frac{\mathrm{d} R_{i}(t)}{\mathrm{d}t}  
    & = &  A \cdot S(\positionCoordinate{i}, t)   \label{eq:droplet_radius}  \\
    \frac{\mathrm{d} \positionCoordinate{i}(t)}{\mathrm{d}t} 
    & = &  \dropletVelocity{i}(t)    \label{eq:position_coordinate}  \\
    \frac{\mathrm{d} \dropletVelocity{i}(t)}{\mathrm{d}t} 
    & = &  \frac{1}{\tau_{p}} [\velocityField(\positionCoordinate{i},t) - \dropletVelocity{i}(t)]  + \gravity.    \label{eq:droplet_velocity} 
\end{eqnarray}
Here, $R_{i}(t)$, $\positionCoordinate{i}(t)$, and $\dropletVelocity{i}(t)$ denote, respectively, the radius, position coordinate, and velocity of the $i$-th droplet; $A$ is a function of temperature and pressure; $S(\positionCoordinate{},t)$ is the supersaturation; $\gravity$ is the gravitational acceleration vector; and $\tau_{p}$ is a measure of the droplet inertial effect.

The condensation rate $C_{d}$ in equations \eqref{eq:temperature}--\eqref{eq:vapor_mixing_ratio} is defined as
\begin{eqnarray}   
    C_{d}(\positionCoordinate{}, t)  & = &
    \frac{4\pi \rho_{l} A}{\rho_{0} a^{3}} \,
    \sum_{i=1}^{n} S(\positionCoordinate{i},t) R_{i}(t) \, ,
    \label{eq:condensation_rate}
\end{eqnarray}
where $a$ is the size of a grid cell, $n$ is the number of droplets in the grid cell, $\rho_{l}$ and $\rho_{0}$ are the densities of water and air, and $R_{i}(t)$ is the radius of the $i$-th droplet.  Note that the condensation rate $C_{d}$ acts as a source (or sink) term for equations \eqref{eq:temperature}--\eqref{eq:vapor_mixing_ratio} and thus couples the particle equations with the fluid dynamics model.  Finally, the full model \eqref{eq:NS}--\eqref{eq:droplet_velocity} is augmented with suitable initial and boundary conditions.
\review{Typical simulations of the PR-DNS model from Gao \textit{et. al} (2018) track evolution of approximately 10 million cloud particles in a physical domain of $0.512 m \times 0.512 m \times 0.512 m$ meshed onto a grid of $256\times 256\times 256$.}

\section{Data Set}
\label{sec:dataset}

As discussed in Section \ref{sec:model_equations}, the PR-DNS model captures the basic fields of fluid dynamics like velocity, temperature, and pressure as well as the cloud physics fields of vapor mixing ratio, supersaturation, and condensation rate.
It also tracks the location and size of cloud particles.
As the first step to model such a complex system, this paper aims to ascertain whether the Navier-Stokes solver governing the velocity field can be replaced by a surrogate model.
To this effect, we select a simpler test case of doubly periodic shear layer where the entire dynamics can be captured by a vorticity or a velocity field.

The doubly periodic shear layer is a well-known test case of two dimensional incompressible flow simulations.
This test has been extensively employed to assess the errors and stability of fluid dynamics models \citep{brown1995performance,atif2022essentially}.
Here, an incompressible fluid of kinematic viscosity $\nu$ is confined in a two dimensional square domain of side $L=1$. 
Periodic boundary conditions are applied in both $x$ and $y$ spatial directions.
The velocity fields at the initial time $t=0$ are given by
\begin{align}
    u_x (y) & =
    \begin{cases}
        U_0 \tanh [(4y-1)/w], \qquad y \leq 1/2\\
        U_0 \tanh [(3-4y)/w], \qquad y > 1/2
    \end{cases}\\
    u_y (x) &= U_0 \delta \sin [2\pi (x+1/4) ],
\end{align}
where $u_x, u_y$ are the velocities in $x,y$ directions respectively. 
The parameters $w$ and $\delta$ control the width of the shear layer and the magnitude of the perturbation.
The Reynolds number ($\rm Re$) defined as ${\rm Re} = U_0 L \,/ \, \nu$ represents the ratio between inertial and viscous forces.
The perturbation causes the shear layers to curl up into two vortices.
We modify the initial condition using random numbers to 
\begin{align}
    u_x  =
    \begin{cases}
        U_0 \, r_1, \qquad y \leq 1/2,\\
        U_0 \, r_2, \qquad y > 1/2,
    \label{eq:initcond1}
    \end{cases}\\
    u_y  = 
    \begin{cases}
        U_0 \, r_3, \qquad x \leq 1/2,\\
        U_0 \, r_4, \qquad x > 1/2,
    \label{eq:initcond2}
    \end{cases}
\end{align}
where $r_1, r_2, r_3, r_4$ are uniformly distributed random numbers between $[-1,1]$. 
This generates two opposite vortices that diffuse as time progresses and convect around the periodic domain depending on their initial locations. 

\begin{figure*}[t]
    \centering
    \includegraphics[width=0.8\textwidth]{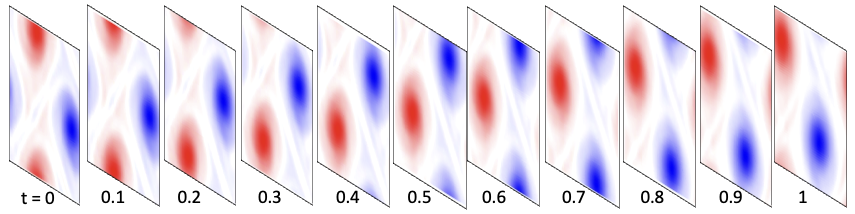}
    \vspace{0.5cm}
    \includegraphics[width=0.8\textwidth]{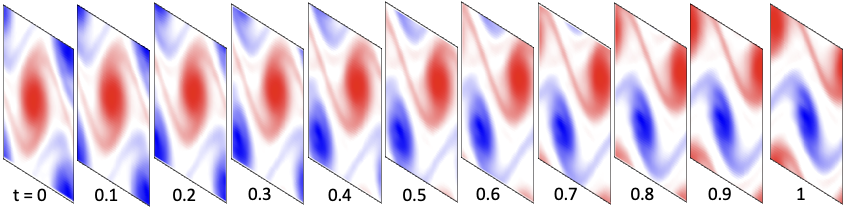}
    \caption{Visualization of vorticity field from two samples of a dataset of 10,000 simulations at ${\rm Re} = 1000$. The red and blue vortices are rotating clockwise and counter-clockwise respectively. The slices at different non-dimensional times $t/t_c$ show the motion of the voritices for unit convection time.}
    \label{fig:dpsl_sample}
\end{figure*}

The data set for the present study corresponds to Reynolds number $1000$.
We solve the Navier-Stokes equations for the velocity field using the entropic lattice Boltzmann method \citep{atif2017essentially,hosseini2023entropic} on a grid of $64 \times 64$ points.
The flow field is first allowed to evolve for $4\, t_c$ (where $t_c = L/U_0$ is the convection time) so that the initial sharp discontinuities vanish.
Thereafter, time is reset to zero and the velocity (${\bf u}$) and vorticity ($\omega_z$) are sampled from time $t=0$ to $t=t_c$ in steps of $0.02 \, t_c$.
Note that the vorticity field $\omega_z(x,y)$ is calculated as the curl of the velocity using
\begin{align}
    \omega_z(x,y) = \nabla \times {\bf u}(x,y),
\end{align}
where $\nabla$ is the nabla operator and the velocity vector ${\bf u}$ has components $u_x, u_y$ for two-dimensional fluid flow.
The dataset consists of fields from 10,000 such simulations. 
\review{The initial condition of the velocity field [Eqns. \eqref{eq:initcond1},\eqref{eq:initcond2}] depend on uniformly distributed random numbers that vary from one sample to another.}
Figure \ref{fig:dpsl_sample} visualizes the two dimensional vorticity field of two such samples.
The two oppositely rotating vortices are visible in the figure.
In general, in an incompressible viscous flow the vortices stretch (absent for two dimensions) and turn under the influence of each other's field and diffuse as the time advances.

\section{Methodology}
\label{sec:methodology}

In recent years, many studies have been conducted to build and evaluate deep learning models as alternatives to traditional PDE solvers for a wide variety of problems in transport phenomenon.
They have been found particularly useful in replacing components of multiscale problems with identifiable computational bottlenecks or in modeling flows for which experimental data is available but constitutive relationships are unknown.
One of the most popular deep learning approach for inverse problems is the physics-informed neural network (PINN) \cite{raissi2019physics}, which incorporates the governing PDEs in the loss function of the minimization problem during training.
PINNs are successful in solving inverse problems for the transport coefficients they are trained on, however, they too sometimes suffer from high computational expense and lack of generalizability.
These lacunae of PINNs are attributed to the fact that they learn mappings between finite dimensional Euclidean spaces.

Neural operators \cite{chen_1995, li2020fourier, deeponet_2021, JMLR:v24:21-1524}, which have gained popularity recently, alleviate the limitations of PINNs.
Particularly, the Fourier neural operators (FNOs) \cite{li2020fourier, JMLR:v24:21-1524} learn mappings between infinite-dimensional function spaces.
\review{A brief description of FNOs is provided in Appendix \ref{app:fno_summary}.}
FNOs have become a popular alternative to the conventional neural networks for a wide range of physical applications like climate modeling \cite{pathak2022fourcastnet}, multiphase flows in porous media \cite{grady2022towards,jiang2023fourier}, and wave propagation \cite{guan2021fourier}.
This popularity is due to their low computational cost, relatively small errors, and the support of features like zero-shot super-resolution for turbulent flows which are limited in other machine learning methods.
Li \textit{et al.} \citep{li2020fourier} proposed two types of FNOs for transient dynamics in two spatial dimensions:
\begin{itemize}
    \item The two dimensional FNO coupled with a recurrent neural network architecture for recurrent propagation in time, which maps the input of a few consecutive time steps to the next chronological time step.
    \item The three dimensional FNO, which learns a mapping in two spatial and one temporal dimensions. This maps an input time window to an entire output time window in a single inference step.
\end{itemize}
In the present study, we choose the three dimensional FNO as we want to predict fields over a long time interval without accumulating errors from step by step temporal predictions.
\review{The Fourier series is truncated at a finite number of lower modes while filtering the higher modes. The FNO model used here has four Fourier layers where width represents the number of nodes in a layer.}
The code is available on Github \citep{li2020fourier} from the original authors of FNO and has also been implemented in NVIDIA's Modulus package \cite{nvidia_modulus}.
We design a simple study to understand the behavior and errors of surrogate models.
As discussed previously, the data set contains velocity and vorticity fields from 10,000 simulations of doubly periodic shear layer corresponding to Reynolds number 1000.
We calculate the errors for different number of samples, input-output training intervals, and time refinements.
The ${\rm L}_2$ errors reported in the following section are calculated at each time step in the prediction interval and averaged over $N=500$ test samples.
\review{We note that the same test set of size 500 is used for all numerical experiments that follow.}
The fields are first normalized via the target data's mean and standard deviation and then the error is calculated as
\begin{align}
{\rm L}_2 \,\,{\rm Error} = \frac{1}{N}\sum_{N} \left[\sqrt{ \frac{1}{n_x n_y} \sum_x \sum_y \left( \phi^{\rm pred}(x,y) - \phi^{\rm true}(x,y) \right)^2 } \right],
\end{align}
where the field $\phi$ is vorticity or velocity, $n_x, n_y$ are the number of grid points in $x$ and $y$ spatial directions, and $N$ is the number of test samples.

\section{Numerical Experiments}\label{sec:results}

\review{The time stepping threshold of numerical schemes for solving partial differential equations are well understood from their stability analysis. However, an analogous threshold for data driven models can not be derived from similar mathematics. The idea behind this section is (a) to experiment with different time steps to understand their relation with the accumulated errors, and (b) to  estimate the amount of data that needs to be generated from PR-DNS model which in itself is a computationally extensive task.}

\subsection{Required number of samples}

\begin{figure}
    \centering
    \includegraphics[width=0.49\textwidth]{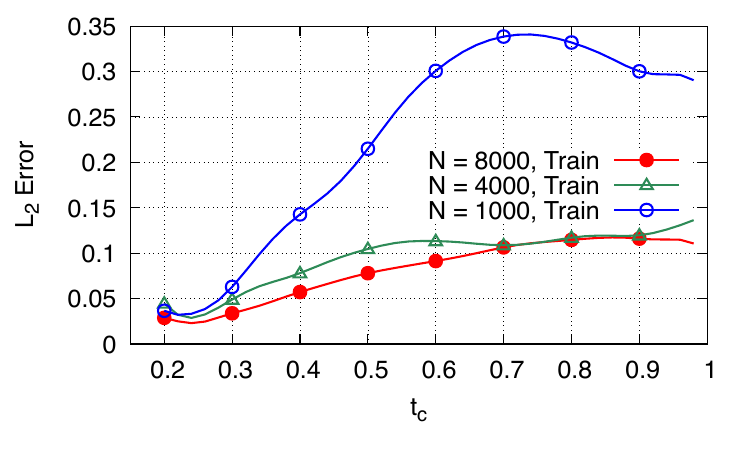}
    \caption{Performance of models trained on $1000$, $4000$, and $8000$ samples as input: ${\rm L}_2$ error calculated on $500$ test samples over the prediction time ($t=0.20 \, t_c$ to $t=0.98 \, t_c$)}
    \label{fig:study1}
\end{figure}

As each simulation of PR-DNS takes substantial time and resources to run and generates a large amount of output data, we would like to assess the smallest number of samples required to train a sufficiently accurate surrogate model. 
To that end, we train three FNO models with the vorticity field using $1000$, $4000$, and $8000$ samples, respectively.
All three models are trained with $10$ time steps ($t=0$ to $t=0.18 \, t_c$) as input and $40$ time steps ($t=0.20 \, t_c$ to $t=0.98 \, t_c$) as output.
We calculate the errors on the same $500$ test samples over the prediction time ($t=0.20 \, t_c$ to $t=0.98 \, t_c$) and plot them in Fig. \ref{fig:study1}.

It is observed from Fig. \ref{fig:study1} that the errors of the model trained with $1000$ samples grows faster with respect to time than the other two models. 
The errors for models trained with $4000$ and $8000$ samples do not have significant difference in terms of prediction errors. 
Thus, in the subsequent tests we will only evaluate models trained with $1000$ and $4000$ samples as this will save significant training data generation costs for the PR-DNS simulations.

\subsection{Choosing the hyperparameters}

Similar to most machine learning models, the Fourier neural operator's performance is dependent on the choice of the so-called hyperparameters.
The hyperparameters here consist of the number of mode, neural net width, learning rate, batch size, scheduler step size, scheduler $\gamma$, etc.
In this section, we calculate the training mean squared error (MSE) with $4000$ training samples under a few batch sizes, widths, scheduler steps, and scheduler $\gamma$s,learning rate, and number of modes. 
Fig. \ref{fig:study3} shows that the MSE with a width 10 is larger than those using other hyperparameter settings (due to its use of smaller neural nets) and \review{with four modes (due to inadequate number of Fourier modes to represent a function)}.
The training time for widths 20 and 40 are similar.
As the width of 40 requires more floating point operations, we select 20 as the default width for the remaining investigations.

\begin{figure}
    \centering
    \includegraphics[width=0.49\textwidth]{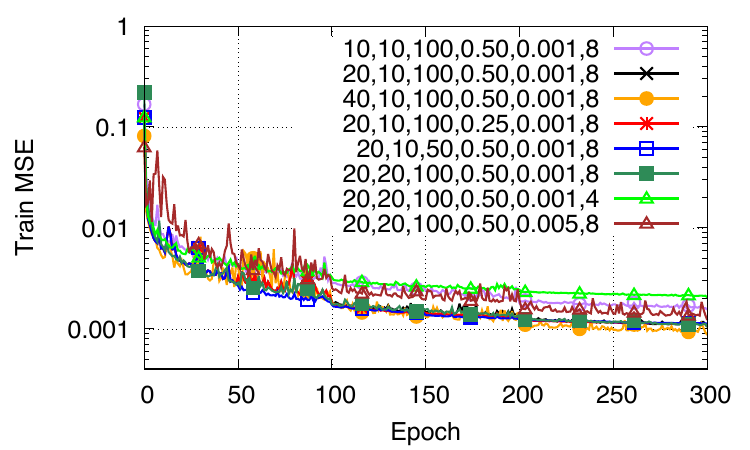}
    \caption{Mean squared error for models trained with 4000 input samples and different hyperparameter settings. The order of hyperparameters from left to right is width, batch size, scheduler step, scheduler $\gamma$, \review{learning rate, and number of modes} respectively.}
    \label{fig:study3}
\end{figure}

\subsection{Varying input and output time windows}

Each sample of doubly periodic shear layer consists of $51$ time snapshots spreading uniformly over unit convection time in steps of $0.02$ convection times. 
As the objective of the study is to accelerate Navier-Stokes solvers, a model with maximum output and minimum input windows is more beneficial computationally.
In the meantime, a surrogate model with low prediction errors for longer output times is preferential.

\review{The data set contains 51 snapshots sampled over unit convection time, which could be split into input and output windows for FNOs in many ways, however, we chose (a) 10 input, 40 output, (b) 20 input, 30 output, and (c) 10 inputs, 10 outputs to understand the trends in their respective errors. }
Thus, we test the error for three input-output time windows: 
\begin{itemize}
    \item input $t= 0$ to $t=0.18 \, t_c$ (10 snapshots); output $t=0.20 \, t_c$ to $t=0.38 \, t_c$ (10 snapshots)
    \item input $t= 0$ to $t=0.18 \, t_c$ (10 snapshots); output $t=0.20 \, t_c$ to $t=0.98 \, t_c$ (40 snapshots)
    \item  input $t= 0$ to $t=0.38 \, t_c$ (20 snapshots); output $t=0.40 \, t_c$ to $t=0.98 \, t_c$ (30 snapshots).
\end{itemize}
This time refinement and input-output split with acceptable errors is expected to depend on the flow conditions too.
For Reynolds number $1000$ we find that ${\rm L}_2$ errors are the lowest for 10 inputs and 10 outputs (see Fig. \ref{fig:study3}).
The plot also shows that the surrogate models accumulate errors for longer prediction times -- a behaviour which is expected to restrict the longest possible prediction times for surrogate models.
\review{This experiment reveals that the errors grow if the input windows are much smaller than the output windows. Thus a model with fewer input snapshots and more output snapshots is expected to perform poorly in comparison to similar number of input output snapshots.}

\begin{figure}
    \centering
    \includegraphics[width=0.49\textwidth]{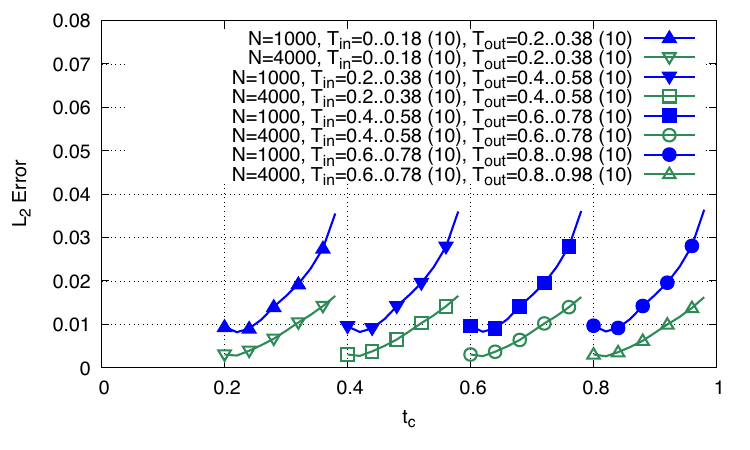}
    \includegraphics[width=0.49\textwidth]{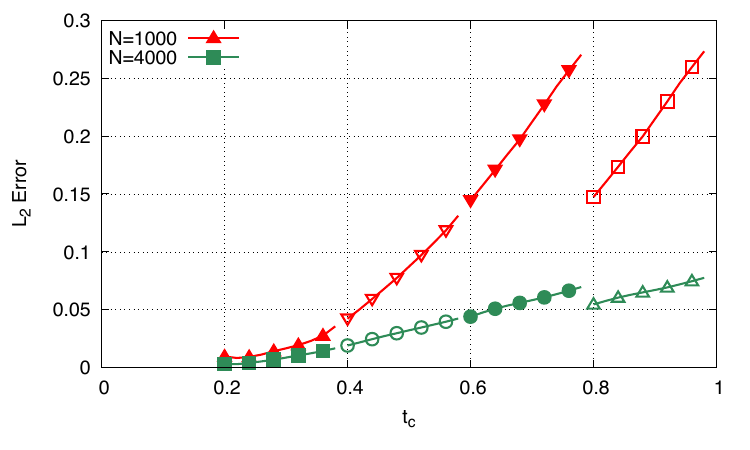}
    \caption{(Top) ${\rm L}_2$ errors with the ground-truth as an input, (bottom) ${\rm L}_2$ errors for the predictions with preceding $0.2 \, t_c$ window as in input. \review{The numbers in the parentheses indicate the number of snapshots in the corresponding data set.}}
    \label{fig:l2_err_vs_time_diff_in}
\end{figure}

Next, we evaluate the errors for input-output time windows different from those for which the model was trained on.
We train the model for $t=0$ to $t=0.18 \, t_c$ as input and $t=0.20 \, t_c$ to $t=0.38 \, t_c$ as output.
Once the model is obtained, we change the input-output time windows to 
\begin{itemize}
    \item input $t=0.20 \, t_c$ to $t=0.38 \, t_c$, output $t=0.40 \, t_c$ to $t=0.58 \, t_c$
    \item input $t=0.40 \, t_c$ to $t=0.58 \, t_c$, output $t=0.40 \, t_c$ to $t=0.58 \, t_c$
    \item input $t=0.40 \, t_c$ to $t=0.58 \, t_c$, output $t=0.60 \, t_c$ to $t=0.78 \, t_c$.
\end{itemize}
Thus, we are essentially computing errors for input-output windows moved ahead in time by $0.2 \, t_c$.
In Fig. \ref{fig:l2_err_vs_time_diff_in}, at the top we plot ${\rm L}_2$ errors with the ground-truth as an input, whereas the bottom figure takes the preceding $0.2 \, t_c$ interval's output as its input. 
It can be seen that when the input is the ground truth the errors remain at relatively low levels.
However while using previous prediction results as input, errors grow more rapidly.
This is expected as inaccurate input amplifies output errors for both numerical simulations and ML models.

\begin{figure}
    \centering
    \includegraphics[width=0.49\textwidth]{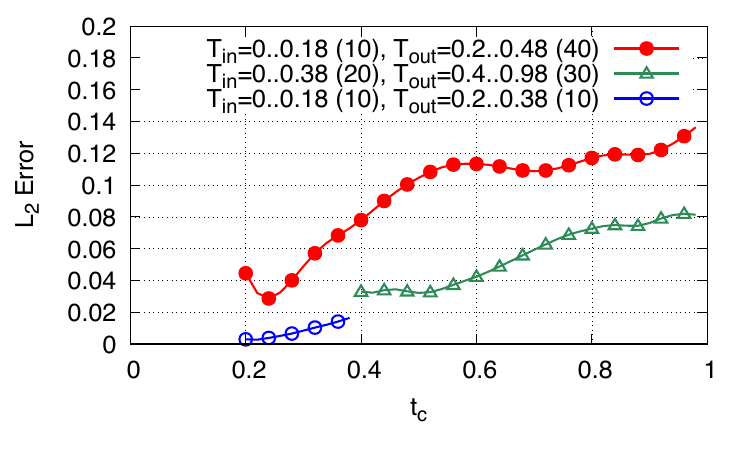}
    \caption{${\rm L}_2$ errors with different input and output time windows for models trained with $4000$ samples. \review{The numbers in the parentheses indicate the number of snapshots in the corresponding data set.}}
    \label{fig:study3}
\end{figure}

\subsection{Varying the time refinement}

\begin{figure}
    \centering
    \includegraphics[width=0.49\textwidth]{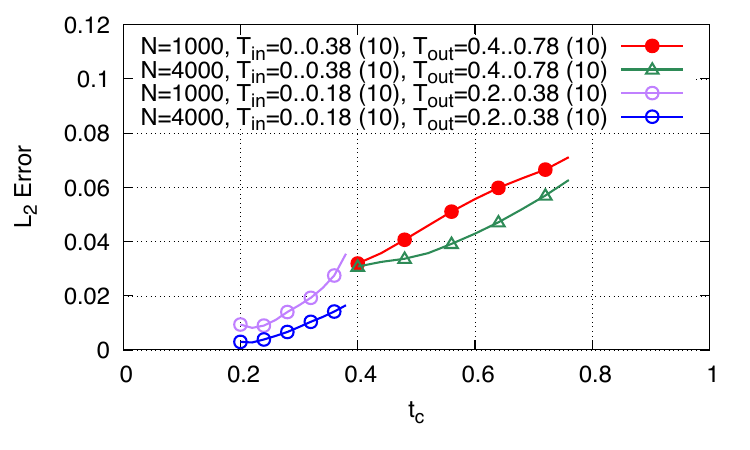}
    \caption{Varying the time refinement of input-output windows for different sample sizes. \review{The numbers in the parentheses indicate the number of snapshots in the corresponding data set.}}
    \label{fig:study4}
\end{figure}

The maximum time step per iteration for classical PDE solvers is typically dependent on convergence and numerical stability. For instance, the Courant-Levy-Friedrichs condition dictates the time step of explicit time integration schemes.
In contrast, the time steps of ML-based surrogate models is not bound to the time step of PDE solvers employed to generate data.
Thus in this section we try to understand the maximum permissible time steps of FNO.
We compare ${\rm L}_2$ errors for two time refinements of $0.02 \, t_c$ and $0.04 \, t_c$, with the input and output time window sizes both fixed to 10. 
For the time refinement of $0.02 \, t_c$, the input window contains 10 time steps from $t=0$ to $t=0.18 \, t_c$ and the output contains 10 time steps from $t=0.20 \, t_c$ to $t=0.38 \, t_c$.
For the time refinement of $0.04 \, t_c$, the input window contains 10 time steps from $t=0$ to $t=0.36 \, t_c$ and the output contains 10 time steps from $t=0.40 \, t_c$ to $t=0.76 \, t_c$.
It is obvious that the latter case is able to predict a longer time window thereby saving substantial computational cost of high-fidelity PR-DNS. 
However, it is important to ensure the errors remain within acceptable limits.
Note that the computational cost of numerical simulation is the same for both the cases below ($0.4 \, t_c$), however this study serves to answer which of the following options is better in terms of accuracy:
\begin{itemize}
    \item Run PR-DNS from $t=0$ to $t=0.18 \, t_c$ then use FNO to predict fields from $t=0.20 \, t_c$ to $t=0.38 \, t_c$. This will be later integrated in a workflow that runs PR-DNS again from $t=0.40 \, t_c$ to $t=0.58 \, t_c$ then uses FNO to predict fields from $t=0.60 \, t_c$ to $t=0.78 \, t_c$.
    \item Run PR-DNS from $t=0$ to $t=0.38 \, t_c$ then use FNO to predict fields from $t=0.40 \, t_c$ to $t=0.78 \, t_c$.
\end{itemize}
The computational cost of the first approach is slightly higher as one needs to perform two FNO predictions, however, it should be noted that the cost of prediction from FNO is only a small fraction of the entire workflow. 
From Fig. \ref{fig:study4} it is seen that the first option accumulates an ${\rm L}_2$ error of $0.018$ till $t=0.18\,t_c$ whereas than the second option accumulates an ${\rm L}_2$ error of $0.064$ till $t=0.78\,t_c$.
If the first option is integrated in a workflow that runs PR-DNS again from $t=0.40 \, t_c$ to $t=0.58 \, t_c$ and performs an FNO prediction from $t=0.60 \, t_c$ to $t=0.78 \, t_c$,
the FNO prediction can be reasonably assumed to accumulate similar ${\rm L}_2$ errors.
Thus, the net error at $t=0.78 \, t_c$ would be around $0.036$ which is smaller than the first approach's $0.064$. 
Therefore, the second approach with smaller intervals but higher time refinement seems to be the better one.

\subsection{Vorticity versus velocity formulation}

\begin{figure}
    \centering
    \includegraphics[width=0.49\textwidth]{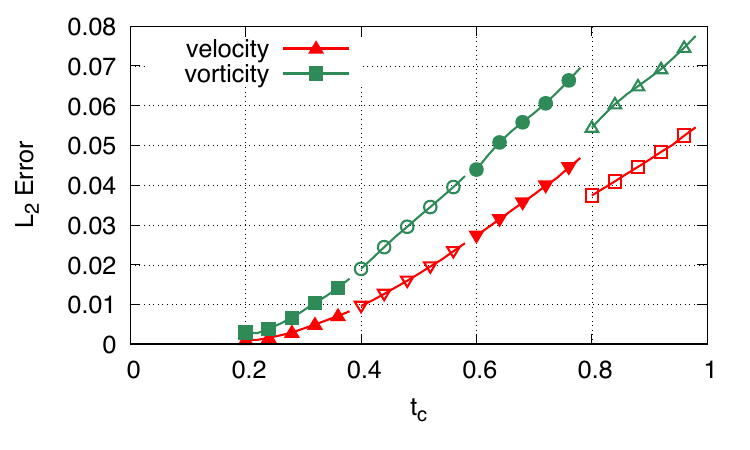}
    \caption{Comparison of ${\rm L}_2$ errors on 500 test samples: vorticity data set versus velocity data set for 4000 training samples. }
    \label{fig:study5}
\end{figure}

The previous experiments are conducted with vorticity field as input and output data.
This approach is advantageous for building surrogate models for incompressible two-dimensional flows because
\begin{itemize}
    \item One only has to deal with one variable i.e., the non-zero component of vorticity vector ${\bm{\omega}} = (0,0,\omega_z)$.
    \item The continuity equation is identically satisfied in the vorticity formulation as one can define a streamfunction $\psi$ such that $\omega_z = - \nabla^2 \psi$ which automatically produces a divergence-free velocity field.
\end{itemize}
It should be noted that the full-fledged PR-DNS model solves the velocity formulation of the three-dimensional Navier-Stokes equations.
Thus, it is important to assess the type and magnitude of errors when dealing with velocity data sets.

In this section, we calculate ${\rm L}_2$ errors when the model training and testing is performed with velocity fields and compare them with the vorticity errors.
It can be observed from Fig. \ref{fig:study5} that the ${\rm L}_2$ error when using the velocity data set is lower than that of vorticity.
Further investigation is required to find the reason for this behavior but it could be an artifact of availability of twice the volume of training data, i.e., two velocity fields $u_x, u_y$ are available as as opposed to a single field of $\omega_z$ in the vorticity formulation.
However, on the flip side a prediction from velocity data set does not guarantee a divergence-free velocity field.

\begin{figure*}
    \centering
    \includegraphics[width=0.98\textwidth]{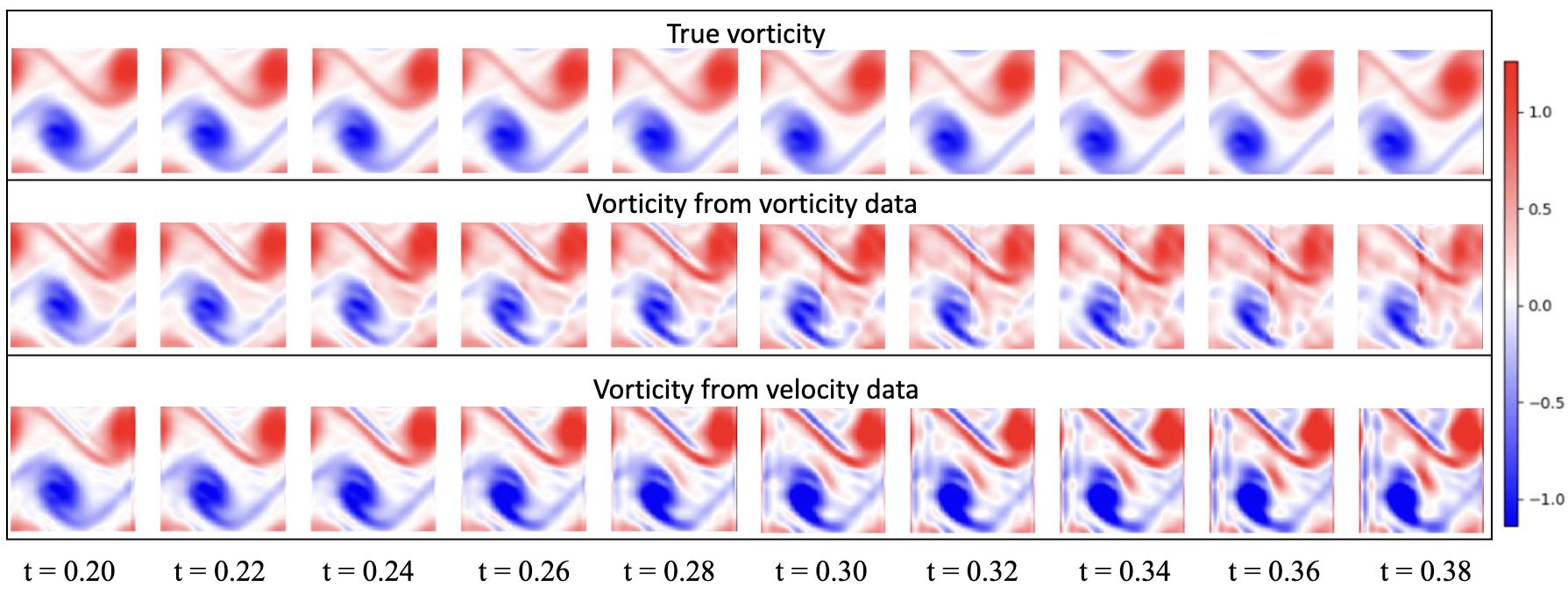}
    \caption{Comparison of vorticity fields of a test sample from training on velocity versus vorticity data. The vorticity field for the velocity data set is found by processing the velocity data.}
    \label{fig:study5p1}
\end{figure*}

\begin{figure*}
    \includegraphics[width=0.98\textwidth]{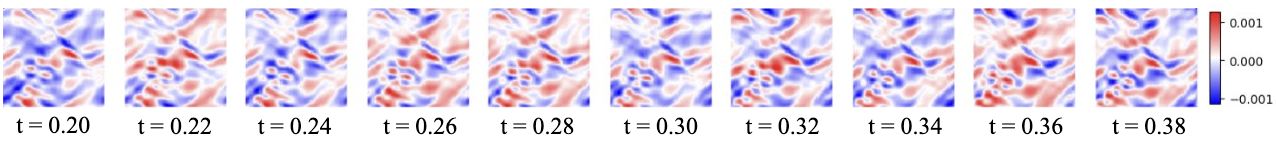}
    \caption{Visualization of the deviation from divergence-free condition for fields obtained from training on velocity data set.}
    \label{fig:study5p2}
\end{figure*}

We visualize the vorticity fields obtained from various experiments in Fig. \ref{fig:study5p1}.
The true vorticity field at the top is the target ground truth.
The middle plot is the vorticity field obtained from the model trained on vorticity data set.
The bottom plot is the vorticity field estimated by processing the velocity data obtained from the model trained on the velocity data set.
It can be seen that the vortices in the target field convect away from the top right corner without any appreciable structural difference.
However, a loss in the braids of the vortices is observed in the middle plot in addition to a noisy vorticity field.
The velocity data set performs better at preserving the braids but shows noise levels similar to the middle plot.
We find that the errors in the continuity equation are of $\mathcal{O}(10^{-3})$ (see Fig. \ref{fig:study5p2}) when using velocity data set.
This will be addressed in the future by incorporating the continuity equation in the loss function.

\section{Conclusions and Future Directions}\label{sec:conclusion}

We have studied the performance of FNO for doubly periodic shear layer at a relatively mild Reynolds number of 1000.
We have found that the lowest prediction errors are obtained when the input-output time intervals are smaller than 0.2 convection times.
We have also seen that the velocity data set performs slightly better than the vorticity data, but, the velocity field obtained is not entirely divergence-free.
Unlike traditional numerical methods at under-resolved grid refinements, FNOs do not give rise to any spurious nonphysical vortices. 
\review{Neglecting the model training cost, the velocity prediction from FNO takes a fraction of the time compared to the numerical PDE solver. For example, the inference from FNO takes 0.21 seconds on Nvidia V100 GPU compared to 0.65 seconds for LBM solver on AMD EPYC 7402P CPU for a mesh of 64x64 over $0.2 t_c$. Thus, by using FNO as the surrogate model for velocity fields we save the cost of the numerical PDE solver which varies from $70-90\%$ (depending on grid size and number of cores used) of the total cost.}

The data from our full-fledged PR-DNS model is three dimensional and corresponds to a Reynolds number of approximately 10,000.
Thus, in the future we plan to extend FNOs to predict three dimensional velocity fields at a higher Reynolds number.
This will require more data in terms of finer temporal and spatial resolutions as well as number of samples.
There are other neural operators such as Clifford \cite{brandstetter2022clifford}, Markov \cite{li2021markov}, and DeepONet \cite{oommen2022learning} that could also be used if they exhibit better performance or yield lower errors.
\review{As this study focuses on only one component of the overall PR-DNS model (i.e., substitution of velocity solver with FNO), it remains to be seen how the other components of the model are impacted by it. However, we note that the errors in FNO's prediction are similar to numerical diffusion of Navier-Stokes solvers at coarse grid resolution in that both manifest as an enhanced diffusion in the vorticity field. This will certainly impact the other variables behavior and will be investigated in future studies.}
Another outstanding question that needs to be answered by future studies is the identification of components of the PR-DNS model other than the Navier-Stokes equations that may benefit if modeled via machine learning methods.  
Although in this work we have evaluated the errors when replacing of the Navier-Stokes solver, our future investigations will assess judicious replacement of thermodynamical governing equations and/or particle propagation with ML-based surrogates as well.
Finally, incorporating governing PDEs in the loss function will lead to smaller errors and is an avenue for further research.

\bmsection*{Acknowledgments}
The authors gratefully acknowledge financial support from the Laboratory Directed Research and Development program at Brookhaven National Laboratory.
We also acknowledge administrative support from the Sustainable Research Pathways (SRP) Program for JY who was SRP student participant at Brookhaven National Lab.

\bmsection*{Financial disclosure}

This work is supported by Laboratory Directed Research and Development program at Brookhaven National Laboratory, which is sponsored by the US Department of Energy, Office of Science, under  Contract Number DE-SC0012704.

\bmsection*{Conflict of interest}

The authors declare no potential conflict of interests.

\bibliography{main}

\bmsection*{Supporting information}

Additional supporting information may be found in the
online version of the article at the publisher’s website.

\appendix
\input{appendix_FNO}





\end{document}

%% file: appendix_FNO.tex
\section{Fourier Neural Operator (FNO) Learning}
\label{app:fno_summary}

\newcommand{\mcirc}{\, \circ \,}

Informally, operator learning deals with the problem of learning mappings between (infinite dimensional) function spaces.  In particular, a neural operator is a (suitably defined) neural network architecture designed to approximate some operator.  Within the context of differential equations, such neural operators are sought as approximations, or surrogates, to the solution operator of differential equations \cite{chen_1995, li2020fourier, deeponet_2021, JMLR:v24:21-1524}. 
Following \cite{JMLR:v24:21-1524}, we provide next a brief description of Fourier Neural Operators (FNOs) for partial differential equations (PDEs), as this is the class of neural operators we adopt in the present study.  For a thorough treatment, the reader is referred to the original publications \cite{li2020fourier,JMLR:v24:21-1524}.

Let $\mathcal{A}$ and $\mathcal{U}$ be two Banach spaces of functions defined, respectively, on bounded domains $\Omega_{d_{a}} \subset \mathbb{R}^{d_{a}}$ and $\Omega_{d_{u}} \subset \mathbb{R}^{d_{u}}$.  Let 
\begin{eqnarray}  \label{eq:operator_G}
    \mathcal{G}: \mathcal{A} \rightarrow \mathcal{U}
\end{eqnarray}
be a (possibly nonlinear) operator mapping functions $a \in \mathcal{A}$ to functions $u \in \mathcal{U}$ and 
\begin{eqnarray}  \label{eq:operator_Gp}
    \mathcal{G}_{\theta}: \mathcal{A} \times \theta \rightarrow \mathcal{U}
\end{eqnarray}
be a map parameterized by $\theta \in \mathbb{R}^{p}$ such that, for some $\theta^{*} \in \mathbb{R}^{p}$, we have that
\begin{eqnarray}  \label{eq:operator_Gopt}
    \mathcal{G}_{\theta^{*}}  \ \approx \  \mathcal{G}
\end{eqnarray}
is a suitable approximation and thus $\mathcal{G}_{\theta^{*}}$ may serve as a useful surrogate to the operator $\mathcal{G}$.  In the case of PDEs, the space $\mathcal{A}$ comprises ``input'' functions, such as initial conditions, coefficients, or forcing terms, that the PDEs depend on, while the solution operator $\mathcal{G}$ maps these inputs to the solution $u \in \mathcal{U}$ satisfying the PDEs with the given inputs.  Since, in practice, solving PDEs numerically is often a time-consuming task, fast and accurate surrogates are frequently needed.

In \cite{JMLR:v24:21-1524}, the authors proposed a neural operator architecture (to be described shortly) and several classes of parameterizations, to serve as surrogates \eqref{eq:operator_Gopt} to operators of interest, such as solution operators of PDEs.  They prove a universal approximation theorem showing that their proposed neural operator, parameterization  classes of which include FNOs, can approximate any given nonlinear continuous operator.  Furthermore, the neural operator proposed in \cite{JMLR:v24:21-1524} is discretization-invariant, meaning that it shares the same model parameters $\theta_{*}$ among different discretizations of the underlying function spaces.  This is advantageous since, upon training (i.e., learning the optimal model parameters $\theta_{*}$), the resulting neural operators can then be used with discretizations of the input space $\mathcal{A}$ and output space $\mathcal{U}$ which are possibly different than the discretizations used when training.

\begin{figure}[htpb]
    \centering
    {{\includegraphics[trim=0.0cm 0.0cm 0.0cm 0.0cm, clip=true, width=0.50\textwidth]{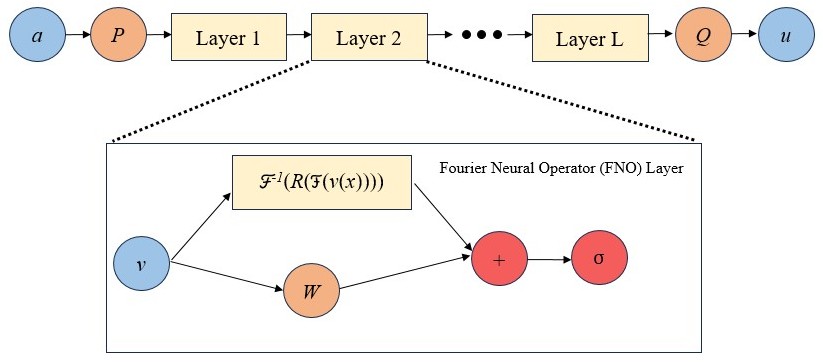}}}
    \caption{Architecture of a Fourier neural operator (FNO).  (Figure adapted from \cite{JMLR:v24:21-1524}.  Permission to reuse material made available under a CC-BY 4.0 license, see https://creativecommons.org/licenses/by/4.0/. Attribution requirements are provided at https://jmlr.org/papers/v24/21-1524.html)}
    \label{fig:fno}
\end{figure}
The architecture of a Fourier neural operator, FNO, is depicted in Figure \ref{fig:fno}.
Here, a FNO $\mathcal{G}_{\theta}: \mathcal{A} \rightarrow \mathcal{U}$ is defined as a composition
\begin{eqnarray}  \label{eq:Gp_fno}
    \mathcal{G}_{\theta} \equiv Q 
        \mcirc \sigma_{L} \mcirc \left( W_{L} + \mathcal{K}_{L} \right) 
        \mcirc \sigma_{L-1} \mcirc \left( W_{L-1} + \mathcal{K}_{L-1} \right)
        \\ \nonumber
        \mcirc  \ \ldots 
        \mcirc \sigma_{1} \mcirc \left( W_{1} + \mathcal{K}_{1} \right) 
        \mcirc P
\end{eqnarray}
of linear operators $\{W_{l} + \mathcal{K}_{l}\}_{l=1}^{L}$ and nonlinear activation functions $\{\sigma_{l}\}_{l=1}^{L}$ for some given number $L$ of layers in the neural network, and where:
\begin{enumerate}
\item First, the input $a \in \mathcal{A}$ is lifted to a higher dimensional representation 
\begin{eqnarray}   \label{eq:lift}
    v_{0}  & = &  P \circ a 
\end{eqnarray}
by a map $P$, often parameterized by a shallow neural network $P_{\theta_{P}}$ with parameters $\theta_{P}$.
\item The above lifting operation is followed by a number $L$ of iterative updates $v_{0} \mapsto v_{1} \mapsto \ldots \mapsto v_{L}$, with
\begin{eqnarray}   \label{eq:fno_layer}
    v_{l}  & = &  \sigma_{l} \circ \left( W_{l} + \mathcal{K}_{l} \right) \circ v_{l-1},   \qquad l = 1, \ldots, L,
\end{eqnarray}
where, for each $l = 1, \ldots, L$, $W_{l}$ is a linear operator parametrized by a neural network $W_{\theta_{W}}$ with parameters $\theta_{W}$ and
\begin{eqnarray}   \label{eq:fno_operator_K}
    \mathcal{K}_{l}  & \equiv &  \mathcal{F}^{-1} \circ R \circ \mathcal{F},  
\end{eqnarray}
where $\mathcal{F}$ denotes the Fourier transform to frequency space and $\mathcal{F}^{-1}$ its inverse (transforming to physical space) and $R$ is an operator filtering frequencies, which is parameterized by a neural network $R_{\theta_{R}}$ with parameters $\theta_{R}$.
\item Finally, the output is 
\begin{eqnarray}   \label{eq:project}
    u  & = &  Q \circ v_{L},  
\end{eqnarray}
where $Q$ is a map that projects $v_{L}$ to $u \in \mathcal{U}$, and is parameterized by a neural network $Q_{\theta_{Q}}$ with parameters $\theta_{Q}$.
\end{enumerate}
The parameters of the FNO $\mathcal{G}_{\theta}$ from \eqref{eq:Gp_fno} are thus given by $\theta = (\theta_{P}, \theta_{W}, \theta_{R}, \theta_{Q})$.